\documentclass[final]{aipproc}
\layoutstyle{6x9}

\SetInternalRegister\hbadness{8000} 

\begin{document}

\title[]{Is the U$_A$(1) symmetry restored at finite temperature or density?}

\author{P. Costa}{address={Centro de F\'{\i}sica Te\'{o}rica,
Departamento de F\'{\i}sica, Universidade, P3004-516 Coimbra, Portugal}}

\iftrue
\author{M. C. Ruivo}{address={Centro de F\'{\i}sica Te\'{o}rica,
Departamento de F\'{\i}sica, Universidade, P3004-516 Coimbra, Portugal}}

\iftrue
\author{C. A. de Sousa}{address={Centro de F\'{\i}sica Te\'{o}rica,
Departamento de F\'{\i}sica, Universidade, P3004-516 Coimbra, Portugal}}

\iftrue
\author{Yu. L. Kalinovsky}{address={Universit\'{e} de Li\`{e}ge, D\'{e}partment de Physique B5, Sart Tilman, B-4000, LIEGE 1, Belgium}}
\fi

\begin{abstract}
We investigate the full U(3)$\otimes$U(3) chiral symmetry restoration, at finite temperature and density, on the basis of the three flavor Nambu-Jona-Lasinio model with the anomaly term given by the 't Hooft interaction. We implement a temperature (density) dependence of the anomaly coefficient motivated by lattice results for the topological susceptibility.
The results suggest that the axial part of the symmetry is restored before the possible restoration  of the full U(3)$\otimes$U(3) chiral symmetry  can occur. 

\end{abstract}
\maketitle


The important role of Quantum Chromodynamics (QCD) at finite temperature and density to describe relevant features of particle physics in the early universe, in neutron stars and in heavy-ion collisions, is nowadays more and more recognized,  bringing together  researches  in lattice QCD, effective models like the Nambu-Jona-Lasinio (NJL) model, and compact star physics calculations.
In fact,  restoration of symmetries and deconfinement are expected to occur  under extreme conditions  that may be achieved in ultra relativistic heavy-ion collisions or in the interior of neutron stars.
An interesting question  is whether  both  chiral SU$(N_f)\otimes$SU$(N_f)$  and axial U$_A$(1) symmetries are restored and which observables could carry information about the possible restorations.

Several studies have been done linking the decrease with temperature of the topological susceptibility, $\chi$, with the restoration of the U$_A$(1) symmetry \cite{lattice}.
There are also preliminary lattice results which indicate the existence of a drop in the behavior of $\chi$ with increasing baryonic density \cite{Alles}.

We perform our calculations in the framework of an extended  SU(3) Nambu--Jona-Lasinio model Lagrangian density that includes the 't Hooft determinant:
\begin{eqnarray}
{\mathcal L\,}&=& \bar q\,(\,i\, {\gamma}.\partial-\,\hat m)\,q
+ \frac{g_S}{2}\, \sum_{a=0}^8[\,{(\,\bar q\,\lambda^a\, q\,)}
^2+{(\,\bar q \,i\,\gamma_5\,\lambda^a\, q\,)}^2\,]  \nonumber \\
&+& g_D\,\{\mbox{det}\,[\bar q\,(1+\gamma_5)\,q] +\mbox{det}
\,[\bar q\,(1-\gamma_5)\,q]\}. 
\end{eqnarray}
By using a standard hadronization procedure, an effective meson action is obtained, leading to gap equations for the constituent quark masses and to meson propagators from which several observables are calculated \cite{costa}.

In the present work we follow the  methodology of \cite{Ohta,Bielich}, and extract the temperature dependence of the anomaly coeficient $g_D$  from the lattice results for the topological susceptibility \cite{lattice}. Other dependences for $g_D$ were studied in \cite{alkofer}.
At temperatures around $T\approx200$ MeV the mass of the light quarks drops 
to the current quark mass, indicating a washed-out crossover. The strange 
quark mass also starts to decrease significantly in this temperature range, 
however even at $T = 400$ MeV it is still 2 times the strange current quark 
mass. So, chiral symmetry shows a slow tendency to get restored in the 
$s$ sector.
In fact, as $m_u=m_d<m_s$, the (sub)group SU(2)$\otimes$SU(2) is a much 
better symmetry of the NJL Lagrangian.
So, the effective restoration of the above symmetry  implies the degeneracy between the  chiral partners $(\pi^0,\sigma)$ and $(a_0,\eta)$ which occurs around $T\approx250$ MeV. At $T\approx350$ MeV both $a_0$ and $\sigma$  mesons become degenerate with the $\pi^0$ and $\eta$ mesons, showing an effective restoration of both chiral and axial symmetries.
So, we  recover the SU(3) chiral partners $(\pi^0,a_0)$ and $(\eta,\sigma)$ 
which are now all degenerated.
However, the $\eta^\prime$ and $f_0$ masses do not yet show a clear tendency 
to converge in the region of temperatures studied \cite{costaUA1}.

Recent calculations on lattice QCD at finite chemical potential  motivates 
also the study of the restoration of the U$_A$(1) symmetry at finite 
density. Since there are no firmly lattice results for the density dependence of $\chi$, to be used as input, we have to extrapolate from our previous results for the finite temperature case and proceed by analogy.
Here we present an example (see Fig. 1, left panel) where we consider quark matter 
simulating "neutron" matter. This "neutron" matter is in $\beta$--equilibrium with charge neutrality, and undergoes a first order phase transition \cite{costa}.
To begin with, we calculate the mixing angles for scalar and pseudoscalar mesons, $\theta_S$ and $\theta_P$, respectively. We observe that $\theta_S$ 
starts at $16^{\circ}$ and increases up to the ideal mixing angle 
$35.264^{\circ}$. A different behavior is found for the angle $\theta_P$ 
that changes sign at $\rho_B\approx4\rho_0$. In fact, it  starts at $-5.8^{\circ}$ and goes to the ideal mixing angle $35.264^{\circ}$, leading to a change of identity between $\eta$ and $\eta'$. We 
think this result might be a useful contribution for the understanding of 
the somewhat controversial question: under extreme conditions will the pion 
degenerate with $\eta$ or $\eta'$? We found that  the change of sign and the 
corresponding change of identity between $\eta$ and $\eta'$, effects that we 
do not observe in the finite temperature case, is related to the small 
fraction of  the strange quarks that only appear in the medium for $\rho_B\approx4\rho_0$ \cite{costa}.

The meson masses, as function of the density, are plotted in Fig. 1, right panel. The results for constant $g_D$ are also presented (middle panel) for 
comparison purposes.
The SU(2) chiral partners ($\pi^0,\sigma$) are  bound states and become 
degenerated at $\rho_B=3\rho_0$.
With respect to the SU(2) chiral partners ($\eta,a_0$), the $a_0$ meson is 
always a purely non strange quark system. For $\rho_B<0.8\rho_0$ $a_0$ is 
above the continuum and, when $\rho_B\geq0.8\rho_0$, $a_0$ becomes a bound 
state. At $\rho_B = 0$, the $\eta$ has a strange component and, as the 
density increases, $\eta$ becomes degenerated with $a_0$ at $4.0\rho_0\leq\rho_B\leq4.8\rho_0$ as expected. In this range of densities 
($\eta,a_0$) and ($\pi^0,\sigma$) are all degenerated. Suddenly the $\eta$ 
mass separates from the others becoming a purely strange state. This is due 
to the behavior of  $\theta_P$ that changes the sign and goes to $35.264^{\circ}$ at $\rho_B\approx4.8\rho_0$. On the other hand, the 
$\eta'$, that starts as an unbounded state and becomes bounded at 
$\rho_B>3.0\rho_0$, turns into a purely light quark system and degenerates 
with $\pi^0$, $\sigma$ and $a_0$ mesons.
Taking into account the presented arguments, we conclude that the U$_A$(1) 
symmetry is effectively restored at $\rho_B>4\rho_0$ \cite{costaUA1}.
In fact, the U$_A$(1) violating quantities show a tendency to vanish, which means that the four meson masses are degenerated and the topological susceptibility goes to zero. Without the restoration of the axial symmetry, the $a_0$ ($\sigma$) mass was moved upwards and never met the $\pi^0$ ($\eta^\prime$ ) mass as can be seen in Fig. 1, middle panel. We remember that the determinant term acts in an opposite way for the scalar and pseudoscalar mesons.

In summary, we have  implemented  a criterion which combines a lattice-inspired behavior of the topological susceptibility with the 
convergence of appropriate chiral partners to explore effective restoration 
of symmetries.
However, the role of U$_A$(1) symmetry for finite temperature, and mainly 
for finite density media, has not been so far investigated and this question 
is still controversial and not settled yet.
We hope that new studies, especially lattice based and experimental ones, 
can finally clarify it.

\begin{figure}[t]
\hspace{0.3cm}\includegraphics[width=7cm,height=6.7cm]{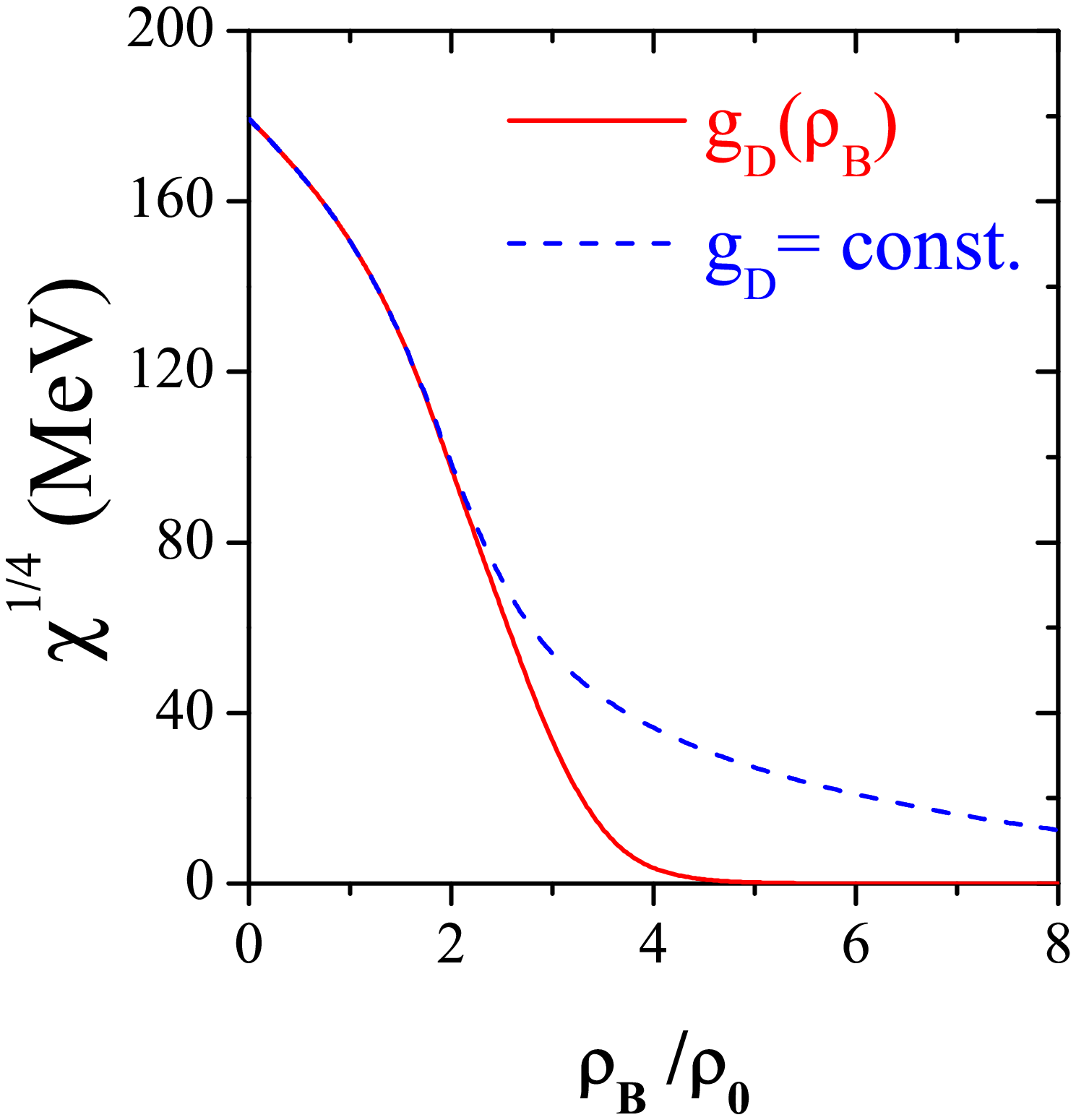}
\hspace{-1cm}\includegraphics[width=10cm,height=7cm]{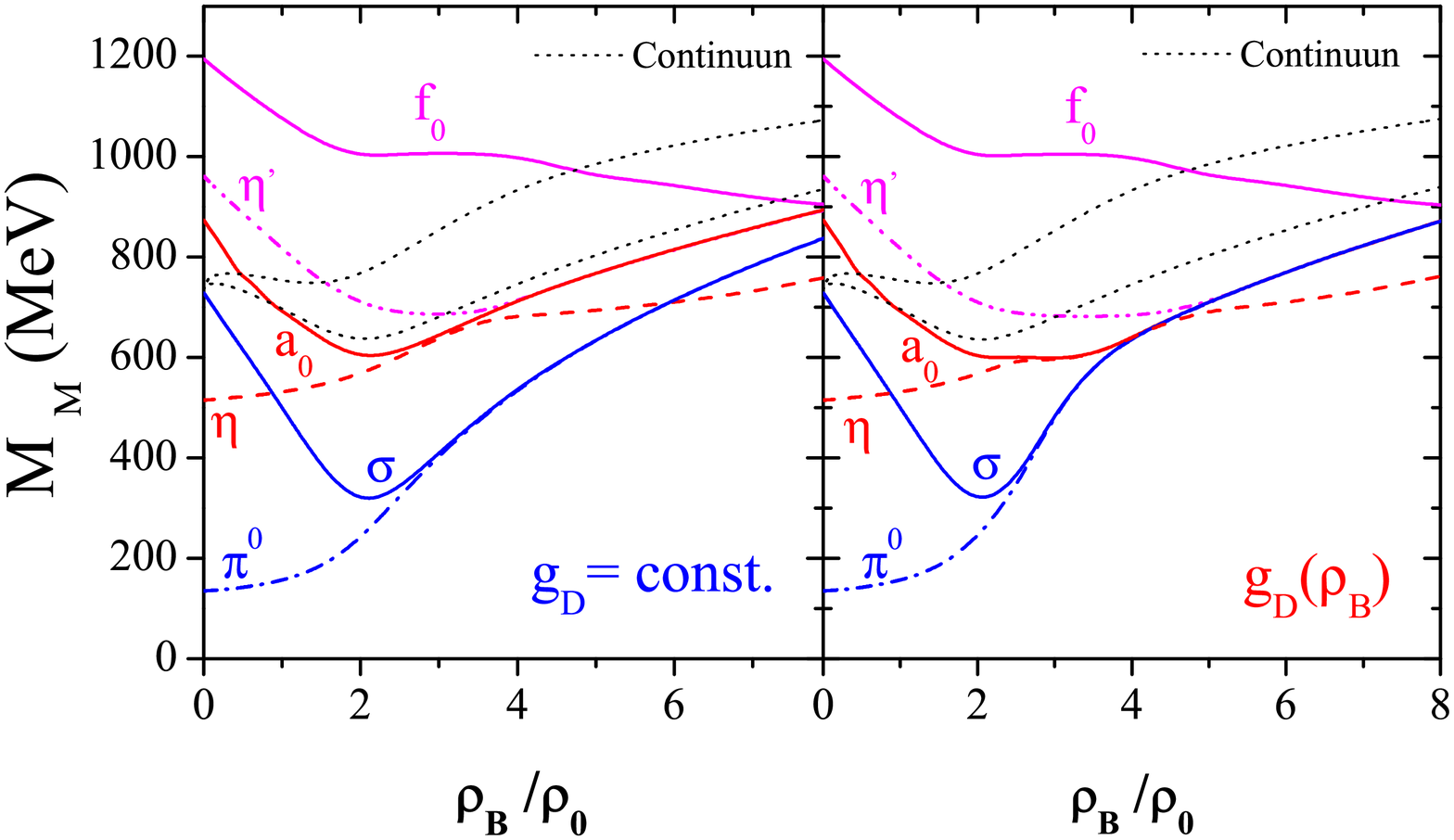}
\caption{Topological susceptibility (left panel): the solid (dashed) line represents our fitting with constant (density dependent) $g_D$. Meson masses, as functions of density, with $g_D$ constant (middle panel) and $g_D (\rho_B)$ (rigth panel). The dotted lines indicate the density dependence of the limits of the Dirac sea continua, defining $q\bar q$ thresholds for $a_0$ and $\eta^\prime$ mesons. }
\label{fig:dens}
\end{figure}

\vspace{0.5cm}
Work supported by grant SFRH/BD/3296/2000 (P. Costa), CFT and by FEDER/FCT under project POCTI/FIS/451/94. 


\end{document}